\documentclass[aps, superscriptaddress, floatfix]{revtex4}

\usepackage{graphicx}

\begin{document}



\newcommand{\Dirac}{\rlap{\hspace{-.5mm} \slash} D}
\newcommand{\sumint}{\rlap{\hspace{-.5mm} $\sum$} \int}
\title{
 A large $N_c$ perspective on the QCD phase diagram}
\author{D.~Toublan}
\affiliation {Physics Department, University of Illinois at Urbana-Champaign,
Urbana, IL 61801}

\date{\today}

\begin{abstract}
The transition between the hadronic
phase and the quark gluon plasma phase at nonzero temperature and quark
chemical potentials is studied within the large-$N_c$ expansion of QCD.
\end{abstract}

\maketitle

\section{Introduction}

The study of QCD at nonzero temperature and quark chemical potentials
is of crucial importance to understand a wide range of different
physical phenomena, from heavy-ion-collision experiments to neutron
stars and cosmology.  This has led to numerous theoretical 
investigations of the phase diagram of QCD at nonzero temperature and
quark-chemical potentials.  

Historically, numerical simulations
using Monte Carlo techniques have been very fruitful for the study
of QCD at nonzero temperature and zero quark chemical potentials.  
However, this approach faces a major
problem at non-zero quark chemical potentials: In general, the fermion
determinant is complex and Monte Carlo techniques do not
work. This is the so-called 'sign problem'. 
There is one exceptional case where there is no  sign problem: 
QCD at non-zero isospin chemical potential and zero
baryon and strangeness chemical potentials.  In this case standard
Monte Carlo techniques can been used. Lattice simulations show that
the QCD phase diagram in this particular case is very rich
\cite{kogut,lattMuI}.   
The sign problem represents a severe challenge for our general
understanding of the QCD phase diagram at nonzero temperature and
quark chemical potentials.  Eventhough there is no general method to
solve the sign problem, recent advances have been made to
circumvent it at small quark chemical potentials
\cite{lattMuB_F&K,lattMuB_Bielefeld,lattMuB_ZH,lattMuB_Maria}.  These
recent studies have concentrated on the transition between the
hadronic phase and the quark-gluon-plasma phase at nonzero baryon
chemical potential and zero isospin and strangeness chemical
potentials, and in particular on the corresponding critical
temperature as a function of quark chemical potentials,
$T_c(\mu_u,\mu_d,\mu_s)$.  

At small chemical potentials, two remarkable
properties of the critical temperature seem to emerge from lattice
studies.  The critical temperature weakly depends on
the chemical potentials, and  $T_c(\mu,\mu,0) \sim
T_c(\mu,-\mu,0)$ \cite{kogut,lattMuB_Bielefeld}.
These properties are rather puzzling,
the second one in particular, since physics at nonzero baryon chemical
potential and zero isospin chemical potential is rather different from
physics at zero baryon chemical potential and nonzero isospin chemical
potential.  In this article we shall use the large-$N_c$ expansion of
QCD to shed light on these properties and show that they naturally
emerge in this context.  We shall also use the large-$N_c$
expansion to show that if there is a first order phase
transition between the hadronic phase and the quark-gluon-plasma phase
at a critical temperature $T_c(\mu,-\mu,\mu_s)\lesssim140$~MeV, 
then there is also a first
order phase transition at a critical temperature 
$T_c(\mu,\mu,\mu_s)$, and vice versa.  This is important since
simulations at $\mu_u=-\mu_d$ can be readily performed using standard
Monte Carlo techniques, whereas simulations at $\mu_u=\mu_d$ suffer
from the sign problem.

\section{Large-$N_c$ expansion}

Since its inception, the large-$N_c$ expansion in QCD
has been a useful tool both conceptually and phenomenologically
\cite{tHooft,largeNc}. It is insightful because it provides us
with a small parameter expansion that distinguishes
classes of Feynman diagrams.  Although the theory has not yet been
solved at leading order in $1/N_c$ in four dimension, 
properties of these classes of Feynman diagrams can be used to determine
relations between some observables. In this article, we shall use the
usual large-$N_c$ expansion \cite{largeNc} augmented by recent
developments on the large-$N_c$ expansion at nonzero temperature and
chemical potential \cite{CohenLargeNmuT}. In the large-$N_c$ expansion, the
leading-order contribution to the pressure is given by planar
diagrams that contain only gluons, and the next-to-leading order
contribution is given by planar diagrams with only one quark loop as a
boundary.  This expansion is still valid at nonzero temperature and
quark chemical potentials, provided that the temperature and the quark
chemical potentials are of order $N_c^0$, and if 
 the different superfluid and superconducting phases
are avoided
\cite{CohenLargeNmuT,cscRev}.   
Following the arguments developed for two flavors in \cite{CohenLargeNmuT},
the large-$N_c$ expansion
leads to the following expression for the pressure in QCD with $N_f$
quark flavors, $f$, in a finite volume
$V$:
\begin{eqnarray}
&p(T,\{m_f,\mu_f\})=\epsilon_0(\{m_f,\mu_f\})+\frac TV \ln Z(T,\{m_f,\mu_f\})
 \\
&p(T,\{m_f,\mu_f\})= N_c^2 \Big( 
p_0(T) + \frac1{N_c} \sum_{f} p_1(T,m_f,\mu_f^2) + 
{\cal  O}(\frac1{N_c^2})  \Big),
\label{pressureNc}
\end{eqnarray}
where $Z(T,\{m_f,\mu_f\})$ is the QCD grand canonical
partition function at nonzero temperature, $T$, for $N_f$ quark
flavors of mass $m_f$ and chemical potential $\mu_f$, and where
$\epsilon_0(\{m_f,\mu_f\})=-\lim_{T\rightarrow0}\frac TV \ln
Z(T,\{m_f,\mu_f\})$ is the vacuum 
energy density.  In the
large-$N_c$ expansion in the equation above,  $p_0(T)$
contains the leading-order diagrams, which are planar diagrams with 
gluons only, and thus depends on $T$ only.  At next-to-leading order
there is no mixing between the different quark flavors. The diagrams
that contribute to 
$p_1(T,m_f,\mu_f^2)$ are planar diagrams that contain gluons and 
only one quark loop as a
boundary, and since the gluons do not mix the quark flavors, the quark
chemical potentials do not mix at next-to-leading order.  
Furthermore, since $Z(T,\{m_f,\mu_f\})=Z(T,\{m_f,-\mu_f\})$ because of
CP, $p_1(T,m_f,\mu_f^2)$ is even in $\mu_f$ \cite{CohenLargeNmuT}.  
The mixing
between the chemical potentials of the different quark
flavors appears only at next-to-next-to-leading order. 

The $U(1)$-axial anomaly could spoil this
property of the diagrammatic expansion of QCD in the large-$N_c$
limit. However, as shown in \cite{anomalyT}, the triangle anomaly does not
depend on the temperature. It is straightforward to generalize the
argument developed in \cite{anomalyT} to show that the anomaly is also
independent of the quark chemical potentials.  Another
way to show that the anomaly does not mix the different quark chemical
potentials is to use chiral perturbation theory in the hadronic phase
at nonzero temperature, zero baryon chemical potential, nonzero
isospin and  strangeness chemical
potentials \cite{GL,KST,SS,KT,STV}. Chiral perturbation theory is a 
low-energy effective theory of QCD solely based on the symmetry of
QCD. It contains only the Goldstone modes due to the spontaneous breaking of
chiral symmetry: $\pi$, $\eta$, and $K$. 
In the large-$N_c$ limit, the chiral symmetry
breaking is different, and the $\eta '$ becomes a Goldstone
boson as well. The large-$N_c$ expansion and the inclusion of the
$\eta '$ have  been implemented in chiral
perturbation theory \cite{GL,Manohar}. The pressure of the hadronic phase can
easily be calculated following \cite{STV}. In the hadronic phase, 
we find that the
temperature and quark chemical potentials enter the pressure and mix 
at ${\cal O}(N_c^0)$ in chiral perturbation theory.
Therefore, in chiral perturbation theory the quark chemical
potentials mix only at 
order $N_c^0$ in the large-$N_c$ expansion in the hadronic phase, 
in complete  agreement
with the diagrammatic argument sketched above and developed in detail
in \cite{CohenLargeNmuT}.  We thus conclude that the anomaly does not
spoil the diagrammatic argument presented in \cite{CohenLargeNmuT}.

\section{Critical temperature}
At finite volume $V$, the separation between the
hadronic phase and the 
quark gluon plasma phase corresponds to a peak in the specific heat
$C_V=\partial \epsilon/\partial T|_V$, where $\epsilon=-p+T \partial
p/\partial T+\sum_f \mu_f \partial p/\partial \mu_f$ is the energy
density.  In the infinite 
volume limit the peak value
of $C_V$  stays finite if there is a crossover, and it diverges 
 if the phase transition is first order or
second order.  The scaling of the
peak value with the volume depends on the nature of the phase transition.
Therefore, the
critical temperature as a function of the quark chemical potential,
$T_c(\mu_u,\mu_d,\mu_s)$, is implicitly given by
\begin{eqnarray}
  \label{critLine}
  \left. \frac{\partial C_V}{\partial T} \right|_{T_c}=0.
\end{eqnarray}
From (\ref{pressureNc}) and (\ref{critLine}) and at zero chemical
potentials, we find that the difference between the critical temperature for
very massive quarks, i.e. pure Yang-Mills theory, $T_c^{\rm YM}$, and
the critical temperature for QCD, $T_c^{\rm QCD}$,  should be of order
$1/N_c$:
\begin{eqnarray}
  \frac{T_c^{\rm QCD}-T_c^{\rm YM}}{T_c^{\rm YM}} =
  {\cal O}(\frac1{N_c}). 
\label{Tcm}
\end{eqnarray}
This is indeed what has been observed on the
lattice for $N_c=3$: $T_c\simeq270$~MeV for pure Yang-Mills \cite{teper}, and
$T_c\simeq175$~MeV for QCD 
with three flavors \cite{karschT,CP-PACSt,MILCt}.  Similarly, the critical
temperatures for pure 
Yang-Mills theories with different number of colors have been computed
on the lattice, and they have been found to differ by 
${\cal  O}(1/N_c^2)$, in agreement with the large-$N_c$ expansion,
\begin{eqnarray}
  \frac{T_c^{\rm YM}(N_c)}{T_\infty^{\rm YM}}=1+\frac{0.76(6)}{N_c^2}+\cdots,
\end{eqnarray}
where $T_\infty^{\rm YM}$ is the critical temperature for pure
Yang-Mills theory when $N_c\rightarrow
\infty$ \cite{teper}. 

At nonzero quark chemical potentials the large-$N_c$ expansion
in (\ref{pressureNc}) implies that for QCD with 
$m_u=m_d$, the critical temperature as a function of quark chemical
potential, $T_c(\mu_u,\mu_d,\mu_s)$, must satisfy the following relation
\begin{eqnarray}
  \label{critT}
  \frac{T_c(\mu,\mu,\mu_s)-T_c(\mu,-\mu,\mu_s)}{T_c(\mu,\mu,\mu_s)}
={\cal  O}(\frac1{N_c^2}).
\end{eqnarray}
Therefore, for  $\mu_s=0$, the critical temperature that
separates the hadronic phase 
and the quark-gluon-plasma phase at nonzero baryon chemical
potential, $\mu_b=(\mu_u+\mu_d)/2$, and zero isospin chemical potential,
$\mu_i=(\mu_u-\mu_d)/2$, differs from the critical temperature at zero
baryon chemical potential and nonzero isospin chemical potential by
$~1/N_c^2$. 
This is in complete agreement with recent results
obtained in numerical lattice simulations
\cite{lattMuB_Bielefeld,kogut}, and in various 
models \cite{qcdMuBMuI_RMT,qcdMuBMuI_NJL,qcdMuBMuI_Ladder}. This is a
useful relation since lattice simulations  
at $\mu_b=0$ and $\mu_i\neq0$ do not suffer from the sign problem present at
nonzero $\mu_b$, and are therefore much easier to perform.

More explicitly and for QCD with three quark flavors with $m_u=m_d\neq
m_s$ in a finite volume, we can perform a 
Taylor expansion of $\partial C_V/\partial T$ around the critical
temperature that separates the hadronic phase from the
quark-gluon-plasma phase at zero chemical potentials, $T_0$. Using
(\ref{pressureNc}), we find that
\begin{eqnarray}
  \label{dCvTaylor}
   \frac{\partial C_V}{\partial T}&=&
  a_1 \frac{T-T_0}{T_0} +
  \frac1{N_c} \Bigg( \Big[ b_0 + b_1 \frac{T-T_0}{T_0} \Big]
  \frac{\mu_u^2+\mu_d^2}{T_0^2} +
  \Big[ b_{s0} + b_{s1} \frac{T-T_0}{T_0} \Big]
  \frac{\mu_s^2}{T_0^2} \Bigg) + \cdots .
\end{eqnarray}
Notice that the coefficients related to $\mu_s$ differ from those
related to $\mu_u$ and $\mu_d$ since $m_u=m_d\neq m_s$. Using
(\ref{dCvTaylor}) to solve (\ref{critLine}), we find that the critical
temperature as a function of the quark chemical potentials is given by
\begin{eqnarray}
  \label{critTTaylor}
  \frac{T_c(\mu_u,\mu_d,\mu_s)}{T_0}=1-
\frac1{N_c} \Bigg( \frac{b_0}{a_1} \frac{\mu_u^2+\mu_d^2}{T_0^2} + 
\frac{b_{s0}}{a_1} \frac{\mu_s^2}{T_0^2}  \Bigg) + \cdots . 
\end{eqnarray}
Therefore, we find that the large-$N_c$ expansion leads to interesting
insight on the critical temperature as a function of quark chemical
potentials.
First the curvature of the critical
temperature for small chemical potential is $1/N_c$ suppressed. 
Second, for a given number of colors $N_c$,
the curvature of the critical temperature as a function of baryon
chemical potential should increase
with the number of flavors.  For $N_f$ degenerate quarks, this increase in the
curvature should be linear in $N_f$ up to ${\cal  O}(1/N_c)$
corrections.  This has been indeed observed in different 
lattice simulations for $N_f=2$,
$3$, and $4$ \cite{lattMuB_ZH,lattMuB_Maria}. 
Third, the leading-order dependence in the quark
chemical potentials is even in $\mu_u$, $\mu_d$, and $\mu_s$
separately, and  does not mix them. The mixing term appears at
next-to-next-to-leading order only. Therefore effects
due to the mixing of the quark 
chemical potentials should be of the order of $1/N_c^2$.  This expression 
for the critical temperature in the large-$N_c$ 
expansion agrees with and simply explains several results that have been
found in numerical simulations and in various models
\cite{lattMuB_F&K, lattMuB_Bielefeld, lattMuB_ZH, lattMuB_Maria, lattMuI,
  kogut, qcdMuBMuI_RMT,qcdMuBMuI_NJL,qcdMuBMuI_Ladder}.

\section{Order of the Phase Transition}
The arguments developed above do not depend on the nature of the phase
transition. This is, however, an important question to address.
Indeed, for QCD with three colors and physical quark masses, 
the separation between the hadronic phase and the quark-gluon
plasma phase at nonzero baryon chemical potential, zero
isospin and strangeness chemical potentials is believed to be a
crossover at low chemical potential and a first order phase transition
at higher chemical potential \cite{ladder,NJL,RMT}.  
Several recent lattice simulations have found
the critical endpoint that corresponds to the end of this first order
phase transition line \cite{lattMuB_F&K, lattMuB_Bielefeld}.  However,
these different simulations yield 
results that significantly differ on the precise location of this
critical endpoint.  

Lattice simulations at nonzero
baryon chemical potential suffer from the sign problem. No such problem
is present at nonzero isospin chemical potential and zero baryon and
strangeness chemical potentials. It is therefore easier to perform
lattice simulations in the latter case. In the section above, we have
shown that the large-$N_c$ expansion leads to  a relation between 
the critical temperature at
$\mu_u=\mu_d$ with the critical temperature at
$\mu_u=-\mu_d$  that is valid at next-to-leading order.
  It is therefore natural to investigate
if such arguments can lead to a relation between the nature of the
phase transition in these two cases.

In the infinite volume limit, if the phase transition is first order,
then there is a latent
heat, $L_h=T_c \; {\rm disc}\;s$, where $s$ is the entropy density. 
The latent heat is related to the
discontinuity of the quark-antiquark condensate,
$\langle\bar{q}q\rangle$, through the Clausius-Clapeyron relation
derived in \cite{leutwylerClausius, gattoClausius} 
\begin{eqnarray}
  \label{Clausius}
  L_h =  \frac{T_c}{\partial T_c/\partial
    m_q|_{\{\mu_f\}}} {\rm disc}\; \langle \bar{q}q \rangle , 
\end{eqnarray}
where $Q|_x$ means that quantity $Q$ is evaluated at constant $x$.
Starting from (\ref{pressureNc}), 
and using the same reasoning that leads to (\ref{critT}),
we find that $\partial T_c/\partial m_q|_{\{\mu_f\}} \sim 1/N_c$.
At zero temperature the large-$N_c$ expansion leads to
$\langle \bar{q}q \rangle_{T=0} \sim {\cal O}(N_c)$.  
At nonzero temperature, $\mu_u=-\mu_d$ and
$\mu_s=0$, based on lattice simulations \cite{lattMuI, kogut} and on chiral
perturbation theory \cite{GerberLeutwyler, STV}, we know that $\langle
\bar{q}q \rangle$ is almost independent of the chemical potential and
of the  temperature in the hadronic phase, provided $T\lesssim
140$~MeV.  For instance in chiral perturbation theory at zero
chemical potentials, it was found
that for $T\lesssim 140$~MeV, $\langle 
\bar{q}q \rangle / \langle \bar{q}q \rangle_{T=0}\gtrsim 80\%$  for
QCD with 
two massive quarks, and with corrections due to massive non-Goldstone
modes taken into account \cite{GerberLeutwyler}.  Thus, from both
lattice simulations 
and chiral perturbation theory, we have that $\langle \bar{q}q
\rangle \sim \langle \bar{q}q \rangle_{T=0}={\cal O}(N_c)$ for $T
\lesssim 140$~MeV in the hadronic phase. Therefore, we 
conclude that if a first order phase transition between the hadronic
phase and the quark-gluon plasma  phase takes place at $T \lesssim
140$~MeV, the corresponding latent heat is  of order
$N_c^2$. This conclusion cannot be reached for higher temperatures
since the value of the quark-antiquark condensate decreases and is no
longer ${\cal O}(N_c)$.

In the large-$N_c$ perspective described above and if there is a first
order phase transition at
$T\lesssim140$~MeV, we can use the same reasoning that led to (\ref{Tcm})
to show that the 
latent heat as a function of quark chemical potentials,
$L_h(\mu_u,\mu_d,\mu_s)$, has to satisfy the following relation 
\begin{eqnarray}
  \label{latentHeat}
  \frac{L_h(\mu,\mu,\mu_s)-L_h(\mu,-\mu,\mu_s)}{L_h(\mu,\mu,\mu_s)}
={\cal  O}(\frac1{N_c^2}), \hspace{1cm} {\rm for} \; T\lesssim140{\rm
  MeV}. 
\end{eqnarray}
In other words, if there is a first order phase transition
between the hadronic phase and the quark-gluon plasma phase 
at a temperature $T_c(\mu,-\mu,\mu_s)\lesssim140$~MeV, 
with a latent heat $L_h = {\cal O}(N_c^2)$, or equivalently ${\rm disc} \;
\langle\bar{q}q \rangle = {\cal O}(N_c)$, 
then, according to the large-$N_c$ expansion, equations (\ref{critT}) and
(\ref{latentHeat}),  there should be a first
order phase transition at a temperature $T_c(\mu,\mu,\mu_s)\sim
T_c(\mu,-\mu,\mu_s)+{\cal O}(1/N_c^2)$. Thus if lattice simulations at
$\mu_u=-\mu_d$ 
were to find a first order phase transition at temperatures below
$\sim140$~MeV, then the $1/N_c$ expansion predicts that there should
also be a first order phase transition at the same $\mu_u=\mu_d$ and 
at the same temperature, up to ${\cal O}(1/N_c^2)$ corrections.

\section{Conclusions}

We have used the large-$N_c$ expansion of QCD to study the phase
transition between the hadronic phase and the quark-gluon plasma phase
at nonzero temperature and quark chemical potentials. We have shown
that the critical temperature depends on the chemical potentials at
next-to-leading order in the large-$N_c$ expansion.  We have also
shown that there are relations between the critical temperature at
nonzero baryon chemical potential and zero isospin chemical potential,
and the critical temperature at nonzero isospin chemical potential and
zero baryon chemical potential. These relations are valid at
next-to-leading order in the large-$N_c$ expansion.
Finally, based on large-$N_c$ arguments, we have shown that 
it should be possible to
determine relatively accurately the position of a first order phase
transition between the hadronic phase and the quark-gluon plasma phase
in the QCD phase diagram at nonzero baryon chemical
potential and zero isospin and strangeness chemical potentials, by
performing ordinary lattice simulations at nonzero isospin chemical
potential and zero baryon and strangeness chemical potentials, as long
as the latent heat is large enough.  This should be a valid
strategy up to temperatures of $\sim140$~MeV 
where the quark-antiquark condensate is not
significantly different from its value at zero temperature.

\begin{acknowledgments}
It is a pleasure to thank P. de Forcrand, J. Kogut, M. P. Lombardo,
A. Manohar, and M. Stephanov for useful discussions. 
\end{acknowledgments}

\end{document}